\documentclass[11pt]{article}
\usepackage{amsfonts}

\font\tengoth=eufm10 \font\sevengoth=eufm7 \font\fivegoth=eufm5
\newfam\gothfam

\textfont\gothfam=\tengoth \scriptfont\gothfam=\sevengoth
  \scriptscriptfont\gothfam=\fivegoth

\newcommand{\sect}[1]{\setcounter{equation}{0}\section{#1}}
\newcommand{\subsect}[1]{\subsection{#1}}

\def\be{\begin{equation}}
\def\ee{\end{equation}}
\def\bea{\begin{eqnarray}}
\def\eea{\end{eqnarray}}

\def\R{{\mathbb R}}
\def\1{\'{\i}}                           
\def\k{\kappa}
\def\>#1{{\bf #1}}                 
\def\d{{\rm d}}

\def\jp{J_+}
\def\jm{J_-}
\def\jj{J_3}
\def\te{\theta}
\def\tes{\phi}
\def\la{\lambda}
\def\rr{r}
\def\tea{r}

\def\om{J}

\def\diag{{\rm diag}}

\def\Sk{{\rm\ \!S}}            
\def\Ck{{\rm\ \!C}}

\def\ff{g}

\begin{document}

\hfill\  
\bigskip

\begin{center} {\Large{\bf{Contractions,  deformations and curvature}}}

\end{center}

\bigskip

\bigskip

\begin{center}
\'Angel~Ballesteros$^{a}$, Francisco J. Herranz$^{a}$, 
  Orlando Ragnisco$^{b}$ and  Mariano Santander$^{c}$
\end{center}

\begin{center} {\it { 
 ${}^a$Departamento de F\1sica, Universidad de Burgos,  E-09001 Burgos, Spain }}\\ e-mail:
 angelb@ubu.es, fjherranz@ubu.es
 \end{center}

 \begin{center} {\it { 
 ${}^b$Dipartimento di Fisica,   Universit\`a di Roma Tre and
 Instituto Nazionale di Fisica Nucleare,  Via Vasca
 Navale 84,  I-00146 Rome, Italy}}\\ e-mail: ragnisco@fis.uniroma3.it
 \end{center}

\begin{center} {\it $^{c}$Departamento de F\'{\i}sica Te\'orica,  Universidad de Valladolid,\\ E--47011
Valladolid, Spain}\\ e-mail: msn@fta.uva.es
\end{center}

\bigskip\bigskip

\begin{abstract}
The role of   curvature in  relation with 
Lie algebra contractions of the pseudo-ortogonal algebras $so(p,q)$ is  fully described by considering 
some associated symmetrical  homogeneous spaces of constant curvature within a Cayley--Klein
framework. We show that a given Lie algebra contraction can  be interpreted geometrically as the
zero-curvature limit of some  underlying   homogeneous space with constant curvature.
In particular, we study in detail  the contraction process for  the three classical
Riemannian spaces (spherical, Euclidean, hyperbolic), three non-relativistic  (Newtonian) 
spacetimes and three relativistic ((anti-)de Sitter and Minkowskian) spacetimes. Next, from
a different perspective, we make use of  quantum deformations of Lie algebras in order to
construct a family of spaces of non-constant  curvature  that can be interpreted as 
deformations of the above nine spaces. In this framework, the quantum deformation parameter
is  identified as the parameter that controls the curvature of such ``quantum" spaces.

\end{abstract}

\bigskip\medskip 

\noindent KEYWORDS:   Lie algebras, quantum groups,  contraction,
curvature, deformation, hyperbolic, de Sitter

\noindent PACS:\quad   02.20.Sv \quad 02.20.Uw  \quad 02.40.Ky 

\bigskip\medskip

\newpage 


\sect{Introduction}

Nowadays, contraction of Lie algebras is a well established 
theory that focuses the interest of both mathematicians and physicists. We recall that  Lie algebra
contractions began to be systematically formulated   from the early works of Segal~\cite{Segal},
In\"on\"u and Wigner~\cite{IWa} and  Saletan~\cite{Saletan}    (see 
also~\cite{Weimar,Pogosyana,montigny} and
references therein). Roughly speaking, the way to obtain a contracted Lie algebra $g'$ from an inital
one  $g$ is to define the generators of $g'$ in terms of those of $g$ in an ``adequate" form by
introducing a contraction parameter $\varepsilon$  in  such a manner that under the limit
$\varepsilon\to 0$ the commutation relations of $g$ reduce to those of $g'$. 

Two well known examples of contraction, related with spaces with arbitrary dimension $N$  are:
(i)~the {\it flat} contraction   that goes from $so(N+1)$ to the Euclidean algebra
$iso(N)$ and (ii)~the {\it non-relativistic} contraction that transforms the Poincar\'e algebra
$iso(N-1,1)$ into the Galilean one $iiso(N-1)$. When looking at the underlying symmetrical homogeneous
spaces associated to the above Lie algebras, one finds that these contractions can geometrically be
interpreted in terms of the vanishing of some {\it (constant) curvature} of such spaces~\cite{Gilmore}. The
former example relates the
  $N$-dimensional ($N$D) spherical space $SO(N+1)/SO(N)$ of constant  curvature $+1/R^2$ ($R$ is
the radius of the sphere) to the flat Euclidean one $ISO(N)/SO(N)$, that is, the limit
$\varepsilon\to 0$ corresponds to $R\to \infty$. The latter contraction  when read as the  ordinary
non-relativistic  limit, relates two {\it flat} spaces (Minkowskian versus Galilean spacetimes) but
alternatively it can also be  naturally interpreted as a contraction starting from the
$2(N-1)$D space of  (time-like) lines
$ISO(N-1,1)/\left( {\mathbb R}\otimes SO(N-1)\right) $  in the flat Minkowskian spacetime
$ISO(N-1,1)/SO(N-1,1)$,  to the flat space of worldlines
$IISO(N-1)/\left({\mathbb R} \otimes SO(N-1)\right) $ in the flat  Galilean spacetime 
$IISO(N-1)/ISO(N-1)$ under the limit $c\to\infty$. The natural curvature of the space of lines  in
Minkowskian spacetime turns out to be non-vanishing and equal to $-1/c^2$, where $c$ is the
relativistic constant, the speed of light. This interpretation of Lie algebra contractions in terms
of zero-curvature limits for   homogeneous spaces can be widely applied for many other cases, which
fully cover the set of possible contractions within the four Cartan families of real semisimple Lie
algebras.

On the other hand, let us consider a {\em  quantum deformation} of the Lie  algebra $g$ endowed with
a Hopf structure~\cite{Abe,Dr,Tjin,CP}, that is, a quantum algebra  $U_z(g)$  which is an algebra of
formal power series in a deformation parameter $z$ ($q={\rm e}^z$) with coefficients in $U(g)$.  In
this case  we know that if a   Lie algebra contraction $g\to g'$ exists   under the limit
$\varepsilon\to 0$,  then this contraction limit can also be  implemented at the deformed level 
$U_z(g)\to U_{z'}(g')$ through a  Lie bialgebra contraction~\cite{LBC}.  The latter  keeps
the same contraction map for the generators while adds some transformation for the contracted
deformation parameter
$z'=z/\varepsilon^n$, where
$n$ is a real number to be fixed for each specific  algebra and contraction. This process is rather
similar to the so called generalized In\"on\"u--Wigner contractions~\cite{Weimar}. By following this
approach, for the two aforementioned      contractions one finds that
$U_z(so(N+1))\to U_{z'}(iso(N))$ and $U_z(iso(N-1,1))\to U_{z'}(iiso(N-1))$ under the limit
$\varepsilon\to 0$.

We stress that a quantum deformation of the Lie algebra $g$ provides an extra   ``quantity"  in the
underlying symmetry structure, the deformation parameter $z$, which can be interpreted in different
ways depending on the specific model under consideration. For instance, as a fundamental scale in
``generalizations" of the special relativity theory~\cite{luki,alu}, as a lattice step in relation
with discretized symmetries~\cite{discrete}, as a coupling constant in $N$-body
problems~\cite{boson}, etc. However, one has to pay the price of loosing the Lie structure and
therefore the corresponding geometric interpretation provided by the associated Lie group and their
homogeneous spaces. Hence, in principle, the geometrical interpretation of ``quantum" contractions in
terms of curvatures seems to be lost.

Nevertheless, if  the Lie algebra contraction  procedure is read in the reverse direction  as a    
Lie algebra   deformation~\cite{nijen} or {\em classical deformation},   a new  interpretation for 
quantum deformations arises in a ``natural" way.  It turns out that the classical deformation $g'\to
g$ (more precisely, from
$U(g')$ to $g$) can be interpreted as the introduction of  a constant curvature in a  formerly   flat
homogeneous space associated to both $g$ and $g'$~\cite{Gilmore,expand}. Such a deformation  process
can be iterated until when one arrives to a semisimple Lie algebra $g$, for which   the   associated
homogeneous spaces (of points, lines, 2-planes, etc.) are endowed with a non-zero constant curvature.
Consequently,  since quantum algebras go beyond   Lie algebras (generalizing them) the above ideas
suggest that a quantum deformation  might also  be understood as the introduction of some kind of
curvature in an appropriate context. In fact, we have recently shown~\cite{plb,sigmaOrlando}  that a
  quantum deformation does indeed introduce curvature   on a certain flat space,  but now this
curvature is generically {\em non-constant} and is governed by the deformation parameter $z$. As a
straightforward consequence, 
 the non-deformed limit $z\to 0$, under which
$U_z(g)\to U(g)\sim g$, can also be understood as a zero-curvature limit, {\it i.e.}  as a
contraction process.

The aim of this paper is to give a brief overview of a global and unified scheme for contractions and 
classical/quantum deformations in relation with curved spaces. Both subjects usually appear as {\em
two} completely separate frameworks in the literature: contractions/classical  deformations   versus
quantum deformations.  For this purpose we shall choose a relevant family of {\em nine} spaces of
constant curvature, namely: the Riemannian (spherical, Euclidean and hyperbolic), semi-Riemannian  
(non-relativistic  oscillating/expanding Newton--Hooke (NH) and Galilean), and pseudo-Riemannian   
(relativistic (anti-)de Sitter and Minkowskian) spaces. Such a family can be described in a common
setting by making use of the so called  orthogonal Cayley--Klein (CK)  or quasi-simple Lie
algebras~\cite{Sommer,Yas,Yaglom,Ros,Gromova,Gromovb,geom,casimir}, all of whose members are
contractions of
$so(N+1)$.

The scheme of the paper is as follows.  In the next Section we recall the structure of  the CK
algebras, we also construct, by using  a  group
theoretical approach, a family of symmetrical homogeneous spaces associated  to each CK algebra, and
next we explicitly  obtain the above nine spaces  in terms of geodesic polar (spherical)
coordinates.  In this way the role of curvature in the context of Lie algebra 
contractions/deformations is highlighted. In Section 3, we start from the non-standard quantum
deformation of
$sl(2,\mathbb R)$~\cite{Ohn}, $U_z(sl(2,\mathbb R))$,  written as a deformed Poisson  algebra and, by
taking three copies of $U_z(sl(2,\mathbb R))$, we are able to construct an infinite  family of 3D
deformed spaces endowed, in general, with a non-constant (scalar) curvature. A further
change of coordinates allows us to introduce spherical coordinates  in such a manner that
we find  that such deformed spaces are just non-constant curvature counterparts  of the
homogeneous spaces described in Section 2. In this way the relationships between curvature
and contraction/quantum deformations can be explicitly analysed. Finally, the last Section
contains some remarks and open problems.


\sect{Contraction, curvature and Lie algebras}


\subsect{Orthogonal CK  algebras}

Let us consider the real Lie algebra $so(N+1)$ whose $\frac 12 N(N+1)$ generators
$\om_{ab}$ $(a,b=0,1,\dots, N$,
$a<b)$ satisfy the   non-vanishing Lie brackets given by  
\be [\om_{ab}, \om_{ac}] =  \om_{bc}  ,\quad [\om_{ab}, \om_{bc}] = -\om_{ac}   ,\quad [\om_{ac},
\om_{bc}] =  \om_{ab}  ,\quad  a<b<c .
\label{aa}  
\ee 
A fine grading group ${\bf {Z}}_2^{\otimes N}$   of $so(N+1)$ is spanned 
by the following $N$ commuting involutive automorphisms $\Theta^{(m)}$ $(m=1,\dots,N)$ of  
(\ref{aa}):
\be
\Theta^{(m)}(J_{ab})=
\left\{
 \begin{array}{rl}
J_{ab},&\mbox{if either $m \le a $ or $b<m$};\cr 
-J_{ab},&\mbox{if   $a<m \le b$} .
\end{array}\right.
\label{aab}
\ee
By applying the  graded contraction theory~\cite{MonPat,MooPat} a large family of  
contracted real Lie algebras can be obtained from
$so(N+1)$; this  depends on $2^N-1$
real contraction parameters~\cite{SolGenGrad} which includes from  the
 simple pseudo-orthogonal algebras $so(p,q)$  (the $B_l$ and $D_l$ Cartan series)  (when all
the contraction parameters are different from zero) up to the Abelian algebra at  the opposite case
(when all the parameters are  equal to zero).   Certainly,  properties associated with the simplicity
of the algebra are lost  at some point beyond the simple algebras in the contraction sequence.
However 
 there exists a particular subset   of contrated Lie algebras 
which are ``close to" to the simple ones~\cite{Grad}, whose members are called   
CK   or quasi-simple orthogonal algebras~\cite{Ros,Gromova,Gromovb,geom}. For
instance, all   the CK algebras share, in any dimension, the same {\em rank} defined  as the number
of (functionally independent) Casimir invariants~\cite{casimir}. These are precisely called CK
algebras  since they  are exactly the family of motion algebras of the geometries of a real space 
with a projective metric in the CK sense
\cite{Sommer,Yas}.

This  orthogonal CK family, here   denoted    $so_{\k}(N+1)$,   depends   on
$N$ real contraction coefficients $\k=(\k_1,\dots,\k_N)$  and, as the essential trait  is the sign of
each $\k_m$,    comprises $3^N$ Lie algebras (up to isomorphisms the number is lesser, as CK algebras
with  different choices of signs for the set     $\k$ may turn out to be isomorphic). The
non-zero commutators   read~\cite{Grad}: 
\be 
[\om_{ab}, \om_{ac}] =  \k_{ab}\om_{bc} , \quad [\om_{ab}, \om_{bc}] = -\om_{ac}  ,\quad [\om_{ac},
\om_{bc}] =  \k_{bc}\om_{ab},  \quad  a<b<c , 
\label{ab}  
\ee without sum over repeated indices and where  the two-index  coefficients $\k_{ab}$ are expressed
in terms of the
$N$ basic  ones through
\be
\k_{ab}=\k_{a+1}\k_{a+2}\cdots\k_b ,\quad a,b=0,1,\dots,N, \quad a<b.
\label{ac}
\ee Each non-zero real coefficient $\k_m$ can be reduced to either $+1$ or $-1$  by a  rescaling of
the Lie generators. The case $\k_m=0$ can be interpreted  as an In\"on\"u--Wigner
contraction~\cite{IWa}, with parameter $\varepsilon_m\to 0$, and defined by the map (cf.\ (\ref{aab})):
\be
\Gamma^{(m)}(J_{ab})=
\left\{
 \begin{array}{rl}
J_{ab},&\mbox{if either $m \le a $ or $b<m$};\cr 
 \varepsilon_m J_{ab},&\mbox{if   $a<m \le b$} .
\end{array}\right.
\label{aac}
\ee

Each involution $\Theta^{(m)}$~(\ref{aab}) provides a Cartan-like decomposition as a   direct sum of
anti-invariant  and invariant subspaces, denoted
${p}^{(m)}$ and ${h}^{(m)}$, respectively:
\be
{{so}}_{\k}(N+1)={p}^{(m)}\oplus {h}^{(m)},
\label{db}
\ee
with the linear sum referring  to the linear structure; Lie commutators fulfil: \be
[{h}^{(m)}, {h}^{(m)}] \subset {h}^{(m)}  , \quad 
[{h}^{(m)}, {p}^{(m)}] \subset {p}^{(m)} , \quad 
[{p}^{(m)}, {p}^{(m)}] \subset {h}^{(m)}   ,
\ee
and thus ${h}^{(m)}$  is always a Lie subalgebra with a
direct sum structure:
\be
{h}^{(m)}={{so}}_{\k_1,\dots,\k_{m-1}}(m)\oplus
{{so}}_{\k_{m+1},\dots,\k_N}(N+1-m),
\label{dc}
\ee
while the vector subspace ${p}^{(m)}$ is generally not a subalgebra.

 The
decomposition (\ref{db}) can be visualized in array form  as follows:
\be
\noindent
\begin{tabular}{cccc|cccc}
$\om_{01} $&$ \om_{02} $& $ \ldots $&$ \om_{0\,m-1} $&$ \om_{0m} $
&$ \om_{0\,m+1}$& $ \ldots $&$ \om_{0N} $  \\
 &$ \om_{12} $& $ \ldots $&$ \om_{1\,m-1} $&$ \om_{1m} $
&$ \om_{1\,m+1}$& $ \ldots $&$ \om_{1N} $  \\
 & & $ \ddots $&$ \vdots $&$ \vdots$
&$  \vdots$& $   $&$ \vdots $  \\
 & &  &$ \om_{m-2\,m-1} $&$ \om_{m-2\,m} $
&$ \om_{m-2\,m+1}$& $ \ldots $&$ \om_{m-2\,N} $  \\
 & &  & &$ \om_{m-1\,m} $
&$ \om_{m-1\,m+1}$& $ \ldots $&$ \om_{m-1\,N} $  \\
\cline{5-8}
 & &  &   \multicolumn{2}{c}{\,}
&$ \om_{m\,m+1}$& $ \ldots $&$ \om_{m\,N} $  \\
 & &  & \multicolumn{2}{c}{\,}
& & $ \ddots $&$ \vdots $  \\
 & &  & \multicolumn{2}{c}{\,}
& &  &$ \om_{N-1\,N} $
\end{tabular}
\label{dia}
\ee
  The subspace
${p}^{(m)}$ is spanned by the $m(N+1-m)$ generators inside the
rectangle; the  left and down
triangles  correspond, in this order, to the subalgebras 
${{so}}_{\k_1,\dots,\k_{m-1}}(m)$ and 
${{so}}_{\k_{m+1},\dots,\k_N}(N+1-m)$ of    ${h}^{(m)}$ (\ref{dc}).

 As some relevant members contained within $so_\k(N+1)$ we point out~\cite{casimir}:

\begin{itemize}

\item   When all
$\k_a\ne 0$ $\forall a$,  $so_\k(N+1)$ is a
(pseudo-)orthogonal algebra $so(p,q)$ ($p+q=N+1$) and
  $(p,q)$ are the number of positive and negative terms  in the invariant   quadratic form with
   matrix 
$(1,\k_{01},\k_{02},\dots,\k_{0N})$.

\item    When $\k_1=0$ we recover the    inhomogeneous algebras 
with   semidirect sum  structure
$$
so_{0,\k_2,\dots,\k_N}(N+1)
\equiv t_N\odot  so_{\k_2,\dots,\k_N}(N)\equiv iso(p,q),  \quad 
p+q=N , 
$$
where the Abelian subalgebra  $t_N$ is spanned by $\langle
\om_{0b};\ b=1,\dots,N\rangle$ and 
$so_{\k_2,\dots,\k_N}(N)$   preserves the
quadratic form with matrix   $\diag (+,\k_{12},\dots,\k_{1N})$.

\item  When $\k_1=\k_2=0$ we get   a ``twice-inhomogeneous" 
pseudo-orthogonal algebra
$$
so_{0,0,\k_3,\dots,\k_N}(N+1) \equiv t_N\odot  \left( t_{N-1}\odot  
so_{\k_3,\dots,\k_N}(N-1)\right)\equiv iiso(p,q),\quad  p+q=N-1,
$$
where the metric of the subalgebra
$so_{\k_3,\dots,\k_N}(N-1)$ is $(1,\k_{23},\k_{24},\dots,\k_{2N})$.

\item  When $\k_a=0$, $a\notin\{ 1,N\}$,    these
contracted algebras can be   described as~\cite{WB}
$$
t_{a(N+1-a)}  \odot (so_{\k_1,\dots,\k_{a-1}}(p,q)\oplus
so_{\k_{a+1},\dots,\k_{N}}(p',q')),\quad p+q=a,\quad p'+q'= N+1-a.
$$

\item   The fully contracted case in the CK family corresponds to
setting all    $\k_a=0$. This is the so called flag algebra 
$so_{0,\dots,0}(N+1)\equiv i\dots iso(1)$ \cite{Ros} such that   $iso(1)\equiv\R$.

\end{itemize}

We recall that the kinematical algebras associated to different models of spacetimes of constant
curvature
\cite{BLLa,BLLb} also belong to these   CK algebras~\cite {SolGenGrad,MonPatTol} and they will be
described in subsection 2.3.


\subsect{Symmetrical homogeneous CK spaces}

If we now consider the CK group $SO_{\k}(N+1)$  with Lie  algebra $so_{\k}(N+1)$ we find that 
each  Lie subalgebra  ${h}^{(m)}$  (\ref{dc}) generates a subgroup $H^{(m)}$  leading  to
the homogeneous coset space denoted by:
\be 
{\cal S}^{(m)} \equiv  SO_{\k}(N+1) \left/  \left( 
SO_{\k_1,\dots,\k_{m-1}}(m)\otimes
SO_{\k_{m+1},\dots,\k_N}(N+1-m) \right) \right. .
\label{dd}
\ee 
The {\em dimension} of ${\cal S}^{(m)}$ is that of ${p}^{(m)}$ (see (\ref{dia})) which is identified
with the tangent space to ${\cal S}^{(m)}$ at the origin:
\be
 \mbox{dim}({\cal S}^{(m)})=m(N+1-m) .
\label{de}
\ee
Then ${\cal S}^{(m)}$ is  a symmetrical homogeneous space (associated to the involution (\ref{aab})),
and there are
  $N$ such symmetrical homogeneous spaces ${\cal S}^{(m)}$ $(m=1,\dots,N)$ for each CK group 
$SO_{\k}(N+1)$. 

Notice that although some Lie algebras in the CK family $so_{\k}(N+1)$ are isomorphic, 
their corresponding sets of $N$ homogeneous spaces are different, as these are determined
not only by $so_{\k}(N+1)$, but also by the subalgebra which will play the role of isotopy
subalgebra of each individual coset space. 
 Furthermore, these $N$ spaces  are not completely unrelated, and it is
possible to reformulate all properties of any given space  in terms of any
other one. In particular,    ${\cal S}^{(2)},
\dots, {\cal S}^{(N)}$ are  usually interpreted in terms of ${\cal S}^{(1)}$, which covers  the
classical Riemannian spaces and spacetimes of constant curvature; 
such an  interpretation  lies in the fact that the subgroups
$H^{(m)}$ ($m=1, 2, \dots, N$) are identified with the isotopy subgroups of
a point ($m=1$), a line ($m=2$),\dots, a hyperplane ($m=N$) in ${\cal
S}^{(1)}$.  Hence, if ${\cal S}^{(1)}$ is taken as {\it the} space, its
elements are called {\it points}, ${\cal S}^{(2)}$ is the space of all
lines in 
${\cal S}^{(1)}$, ${\cal S}^{(3)}$ is the space of all 2-planes in 
 ${\cal S}^{(1)}$, etc.

\begin{table}[h] {\footnotesize
 \noindent
\caption{{Isotopy subgroup, sectional curvature, dimension and rank of the set of $N$ symmetrical
homogeneous spaces ${\cal S}^{(m)} \equiv  SO_{\k}(N+1)/ H^{(m)}$.}}
\label{table1}
\medskip
\noindent\hfill
$$
\begin{array}{lllc}
\hline
\\ [-6pt]
\mbox {Isotopy subgroup} &\!\! \mbox {Curv.} &\mbox {Dimension} &\mbox {Rank} \\[4pt] 
\hline
\\ [-6pt]
  H^{(1)}   =SO_{\k_2,\dots,\k_N}(N)&\k_1&N&1\\ 
H^{(2)} =SO_{\k_1}(2)\otimes SO_{\k_3,\dots,\k_N}(N-1)
&\k_2&2(N-1)&2\\ 
H^{(3)} =SO_{\k_1,\k_2}(3)\otimes SO_{\k_4,\dots,\k_N}(N-2) &\k_3&3(N-2)&3\\ 
\qquad \vdots &\ \vdots&\ \vdots& \vdots\\
H^{(m)} =SO_{\k_1,\dots,\k_{m-1}}(m)\otimes SO_{\k_{m+1},\dots,\k_N}(N+1-m)
&\k_{m}&m(N+1-m) &{\rm min}\,(m,N+1-m)\\ 
 \qquad \vdots &\  \vdots&\  \vdots& \vdots\\
H^{(N-2)} =SO_{\k_1,\dots,\k_{N-3}}(N-2)\otimes SO_{\k_{N-1},\k_N}(3) &\k_{N-2}& (N-2)3&3\\ 
H^{(N-1)} =SO_{\k_1,\dots,\k_{N-2}}(N-1)\otimes SO_{\k_N}(2) &\k_{N-1}& (N-1)2&2\\ 
H^{(N)} =SO_{\k_1,\dots,\k_{N-1}}(N)  &\k_{N}&N&1\\[4pt]
\hline
\end{array}
$$
\hfill}
\end{table}

We   define the {\it rank} of  the CK space
${\cal S}^{(m)}$ as the number of independent invariants under the action of
the CK group for each generic pair of elements in  
${\cal S}^{(m)}$ (see~\cite{Jordan} for the Euclidean case); such a number 
turns out to be the same for all   ${\cal S}^{(m)}$, so it does not depend on the values of
$\k$: 
\be
 \mbox{rank}({\cal S}^{(m)})=\mbox{min}( m,N+1-m).
\label{df}
\ee
Thus, ${\cal S}^{(1)}$  has a single invariant (the ordinary
distance) associated to each pair of points;  ${\cal S}^{(2)}$  has two invariants for each pair
of lines (a ``stationary angle"   and  a ``distance" between the two lines),  and, in
general, ${\cal S}^{(m)}$ has (\ref{df}) invariants for a pair of $(m-1)$-planes
(these are called collectively ``stationary angles"; the last of these is a single  ``stationary
distance").

  The sectional {\em curvature} of ${\cal S}^{(m)}$ turns out to be {\it essentially}  constant and
equal to $\k_m$ (warning: we do not enter here into the details required to  precise the
``essentially"; let we only mention that in a space of rank equal to $r$ there are flat
$r$-dimensional subspaces; as the sectional curvature must vanish along any plane direction contained
in  such a  subspace it cannot, of course, be constant in the ordinary sense; however the statement is
meaningful with the due qualifications). We display  in table~\ref{table1} all these  results
concerning ${\cal S}^{(m)}$.

So far, we have interpreted each (graded) contraction parameter $\k_m$, appearing in  an ``abstract"
form in the commutation relations of $so_\k(N+1)$ (\ref{ab}), as the constant curvature of the
associated symmetrycal homogeneous space ${\cal S}^{(m)}=SO_\k(N+1)/H^{(m)}$. 
Then when the contraction process is
read in the reverse way, as a classical deformation one, we find that to introduce a non-zero constant
$\k_m$ in the Lie brackets (\ref{ab}) geometrically corresponds to the obtention of
a curved space  ${\cal S}^{(m)}$ with $\k_m\ne 0$ from an initial flat one with $\k_m=0$; notice
that the    ``flat"   and ``non-relativistic"   contraction/deformation examples commented
in the introduction are recovered as  two very particular cases within this framework
for $\k=(0,+\dots,+)\leftrightarrow (+1/R^2,+\dots,+)$ and $(0,0,+\dots,+)\leftrightarrow
(0,-1/c^2,+,\dots,+)$, respectively.
In this sense, we recall that the
usual approach to Lie algebra deformations~\cite{montigny,nijen} introduce non-zero structure
constants in a given Lie algebra  leading to a ``less" Abelian one  by applying  
cohomology techniques (see, e.g.,~\cite{figue} for a complete description of Galilean  deformations).
Thus  by starting from the flag algebra $so_{0,\dots,0}(N+1)$ one could reach the simple
ones $so_\k(N+1)$ with all $\k_m\ne 0$. Nevertheless such an algebraic procedure does not
focus on the underlying homogeneous spaces. In contrast, 
an alternative deformation formalism makes use of
universal enveloping algebras and of their associated  homogeneous spaces~\cite{Gilmore,expand}, in
such a manner that the deformed generators are written as elements of the universal enveloping
algebra to be deformed.    We omit here the details, but we remark that   this approach is
directly related with the CK scheme of homogeneous spaces and,  furthermore,  this suggests some kind
of relationship with quantum algebras, as these are also constructed within universal enveloping
algebras; in fact  we will  establish such a connection in section 3.


\subsect{Riemannian and (non-)relativistic spaces of constant curvature}

 From now on  we assume that $\k_3=\dots =\k_N=+1$ and consider 
the rank-1 $N$D space 
$$
{\cal S}^{(1)}=
SO_{\k_1,\k_2,+,\dots,+}(N+1)/SO_{\k_2,+,\dots,+}(N)\equiv
SO_{\k_1,\k_2}(N+1)/SO_{\k_2}(N)\equiv  {\mathbb S}^{N}_{[\k_1]\k_2},
$$
 with sectional
curvature $\k_1$ and metric with signature determined by $\k_2$ through  the  matrix 
${\rm diag}(+1,\k_{2},
\dots,\k_{2})$. Hence we shall deal with the following nine well known   spaces of constant
curvature:

\begin{itemize}

 \item When $\k_2>0$, ${\mathbb S}^{N}_{[\k_1]+}$   covers the
three classical Riemannian spaces: 
$$
\begin{array}{ll}
{\mbox {Spherical:}}  &\quad{\mathbb S}^{N}_{[+]+}\equiv {\bf S}^N=SO(N+1)/SO(N).\\[4pt]
\mbox {Euclidean:}&\quad {\mathbb
S}^{N}_{[0]+}\equiv{\bf E}^N=ISO(N)/SO(N).\\[4pt]
\mbox {Hyperbolic:}&\quad {\mathbb
S}^{N}_{[-]+}\equiv{\bf H}^N=SO(N,1)/SO(N). 
\end{array}
$$
 Their   curvature can be written as $\k_1=\pm 1/R^2$ where $R$ is the
radius of the space  ($R\to \infty$ for the Euclidean case).

\item
  When $\k_2<0$  we get a Lorentzian metric corresponding to  
 relativistic spacetimes~\cite{BLLa}: 
$$
\begin{array}{ll}
{\mbox {Anti-de Sitter:}}  &\quad {\mathbb S}^{N}_{[+]-}\equiv{\bf
AdS}^{(N-1)+1} =SO(N-1,2)/SO(N-1,1).\\[4pt]
\mbox {Minkowskian:}&\quad {\mathbb S}^{N}_{[0]-}\equiv{\bf M}^{(N-1)+1}
=ISO(N-1,1)/SO(N-1,1).\\[3pt]
\mbox {de Sitter:}&\quad {\mathbb S}^{N}_{[-]-}\equiv{\bf dS}^{(N-1)+1} =SO(N,1)/SO(N-1,1). 
\end{array}
$$
The two contraction parameters can be expressed as 
$\k_1=\pm 1/\tau^2$, where $\tau$ is the (time) universe radius, and $\k_2=-1/c^2$,  where $c$ is the
speed of light.

\item
 The contraction $\k_2=0$ $(c\to \infty)$ gives rise to the  non-relativistic  spacetimes with a
degenerate metric~\cite{BLLa}: 
$$
\!\!\!\!\!\!\!\!\!\!\!\!\!\!\!\!\!\!
\begin{array}{ll}
{\mbox {Oscillating NH:}}  &\  {\mathbb S}^{N}_{[+]0
}\equiv{\bf NH}_+^{(N-1)+1} =T_{2N-2}\odot( SO(2)  \otimes  SO(N-1))/ISO(N-1).\\[4pt]
\mbox {Galilean:}&\  {\mathbb
S}^{N}_{[0]0}\equiv{\bf G}^{(N-1)+1} =IISO(N-1)/ISO(N-1).\\[2pt]
\mbox {Expanding NH:}&\  {\mathbb
S}^{N}_{[-]0}\equiv{\bf NH}_-^{(N-1)+1} =T_{2N-2}\odot(SO(1,1)\otimes  SO(N-1) )/ISO(N-1). 
\end{array}
$$

\end{itemize}

In what follows  we
construct  an explicit model of the   space ${\mathbb S}^{N}_{[\k_1]\k_2}$ in terms of
$(N+1)$ ambient coordinates and of $N$ intrinsic     quantities.  The CK algebra
$so_{\k_1,\k_2}(N+1)$  has a  vector  representation given by the following $(N+1)\times (N+1)$ real
matrices fulfilling (\ref{ab}):
\be
J_{ab}=-\k_{ab}e_{ab}+e_{ba},
\label{ccdd}
\ee where $e_{ab}$ is the matrix 
 with   entries
$(e_{ab})_m^l=\delta_a^l\delta_b^m$. In this realization, any element $X \in so_{\k_1,\k_2}(N+1)$
satisfies the equation:
\be
X^T \mathbb I_{\k}+\mathbb I_{\k} X=0 ,
\qquad \mathbb I_{\k}={\rm diag}(+1,\k_1,\k_1\k_2,\dots,\k_1\k_2) ,
\label{af}
\ee
where $X^T$ is the transpose matrix of $X$. Hence  any element $G\in 
SO_{\k_1,\k_2}(N+1)$ verifies $G^T \mathbb I_{\k}   G=\mathbb I_{\k} $ and
$SO_{\k_1,\k_2}(N+1)$ is a group of isometries of  $\mathbb I_{\k}$ acting on
a linear ambient space
$\mathbb R^{N+1}=(x_0,x_1,\dots,x_N)$ through matrix multiplication. The origin $\cal O$ in ${\mathbb
S}^{N}_{[\k_1]\k_2}$ has $(N+1)$ ambient coordinates ${\cal O} =(1,0,\dots,0)$ which is invariant
under the  subgroup $H^{(1)}=SO_{\k_2}(N)$. The orbit
of ${\cal O}$ corresponds to   ${\mathbb S}^{N}_{[\k_1]\k_2}$ which is
contained in the ``sphere" determined  by $\mathbb I_{\k}$:
\begin{equation}
\Sigma\equiv x_0^2+\k_1 x_1^2+\k_1\k_2 \sum_{j=2}^N x_j^2  =1 .
\label{cd}
\end{equation}
 The CK metric on ${\mathbb
S}^N_{[\k_1]\k_2}$
follows from    the flat ambient metric in $\mathbb R^{N+1}$  in the form
\begin{equation}
{\rm d} s^2_{\rm CK}= \frac {1}{\k_1}
\biggl({\rm d} x_0^2+   \k_1 {\rm d} x_1^2+\k_1\k_2  \sum_{j=2}^N {\rm d} x_j^2 
 \biggr)\biggr|_{\Sigma}.
\label{ce}
\end{equation}
 
 Next we parametrize the $(N+1)$ ambient coordinates $\>x$  of a generic point $\cal P$  in
terms of $N$ intrinsic quantities
$(\tea,\theta,\tes_3,\dots,\tes_N)$ called {\em geodesic polar coordinates}~\cite{conf} on   
 ${\mathbb S}^N_{[\k_1]\k_2}$   through the following action of $N$
one-parametric subgroups of $SO_{\k_1,\k_2}(N+1)$ on  $\cal O$:
\be
   \>x =  \exp(\tes_N J_{N-1\, N})\exp(\tes_{N-1} J_{N-2\, N-1})\dots\exp(\tes_3
J_{23})\,\exp(\theta
J_{12})\,
\exp(\tea J_{01})\, \cal O , 
\label{bba}
\ee
which yields 
 \be 
\begin{array}{l}
x_0 
= \Ck_{\k_1}(\tea),\\
x_1=  \Sk_{\k_1}(\tea)\Ck_{\k_2}(\theta ),\\
x_i 
=\Sk_{\k_1}(\tea) \Sk_{\k_2}(\theta )   \prod_{s=3}^i\sin\tes_s\cos\tes_{i+1}, \\
 x_N =\Sk_{\k_1}(\tea) \Sk_{\k_2}(\theta )\prod_{s=3}^N\sin\tes_s  , 
\end{array}
\label{bbb}
\ee
where $i=2,\dots,N-1$ and  any product $\prod_s^i$ where $s>i$
is assumed to be equal to 1.
The $\k$-trigonometric functions $\Ck_{\k}(x)$ and $\Sk_{\k}(x)$ are 
 defined by~\cite{conf} (here for $\k\in \{\k_1,\k_2\}$):
\be 
\Ck_{\k}(x) 
=\left\{
\begin{array}{ll}
  \cos {\sqrt{\k}\, x}, &\   \k >0, \cr 
  1  ,&\  \k  =0 ,\cr 
\cosh {\sqrt{-\k}\, x} ,&\  \k <0 .
\end{array}\right. \qquad 
\Sk_{\k}(x) 
=\left\{
\begin{array}{ll}
    \frac{1}{\sqrt{\k}} \sin {\sqrt{\k}\, x} ,&\   \k >0, \cr 
  x ,&\  \k  =0 ,\cr 
\frac{1}{\sqrt{-\k}} \sinh {\sqrt{-\k}\, x}, &\  \k <0. 
\end{array}\right.  
\label{ae}
\ee
 
The    (physical) geometrical role of these  coordinates is as follows.
Let us consider
a  (time-like) geodesic  $l_1$    and other $(N-1)$  (space-like)
geodesics  $l_j$ $(j=2,\dots,N)$  in ${\mathbb S}^N_{[\k_1]\k_2}$ which are orthogonal at
the origin $\cal O$  in suh a manner that each translation generator  $J_{0i}$ moves $\cal O$ along
$l_i$. Then, 
\begin{itemize}
\itemsep=0pt
\item The radial coordinate $r$ is the   distance between the point $\cal P$ and the origin   $\cal
O$ measured along the   geodesic $l$ that joins both points. In the  
Riemannian spaces with
$\k_1=\pm 1/R^2$,
$r$ has dimensions of {\it length}, $[r]=[R]$; notice   that   the 
dimensionless  coordinate $r/R$  is usually
taken  instead of $r$ and  $r/R$  is     an ordinary angle~\cite{Pogosyanb}.
 In the   spacetimes with
$\k_1=\pm 1/\tau^2$,  $r$ has dimensions of a time-like length,
$[r]=[\tau]$.

\item The coordinate  $\theta$ is an ordinary angle  in the three Riemannian spaces
($\k_2=+1$),  while it
corresponds to a rapidity in the   spacetimes ($\k_2=-1/c^2$)  with dimensions
$[\theta]=[c]$. For the nine spaces,  $\theta$   parametrizes the orientation of $l$ with respect
to the basic (time-like) geodesic $l_1$.

\item The remaining $(N-2)$ coordinates $\tes_3,\tes_4,\dots,\tes_N$  are ordinary    angles for
the nine spaces and correspond to the  polar angles of $l$ relative
to the reference flag at the origin $\cal O$
 spanned by $\{l_1, l_2\}, \{l_1, l_2,l_3\}, \dots,
\{l_1,\dots,l_{N-1}\}$, respectively.

\end{itemize}

In the   three Riemannian cases $(\tea,\theta,\tes_3,\dots,\tes_N)$ parametrize the complete space, while
in the  relativistic spacetimes these only cover the time-like region limited by the light-cone on which
$\theta\to\infty$. The  flat contraction $\k_1=0$ gives rise to the usual spherical
coordinates in the Euclidean space (with $\k_2=+1$).

By introducing  (\ref{bbb})  in   (\ref{ce}),   we obtain the
 CK metric in ${\mathbb
S}^N_{[\k_1]\k_2}$ expressed in geodesic polar coordinates:
\be
 \d s^2_{\rm CK}   =\d\tea^2+\k_2\Sk^2_{\k_1}(\tea) \left\{  \d\theta^2+
 \Sk^2_{\k_2}(\theta)  \sum_{i=3}^{N} 
\left( \prod_{s={3}}^{i-1} \sin^2 \tes_s \right) \d\tes_i^2 \right\} .
 \label{bbc}
\ee
The sectional  $K_{ij}$ and the  scalar  $K$ curvatures are $K_{ij}=\k_1$
and $K=N(N-1)\k_1$; in this rank-one case, the sectional curvature along any 2D-direction  equals
precisely $\k_1$ and hence is actually constant in the literal sense.

As an example, which is   also necessary for our further development  in relation with quantum
deformations, we display in table
\ref{table2} these results for $N=3$.

\begin{table}[t] {\footnotesize
 \noindent
\caption{{Metric, sectional and scalar curvatures of the nine 3D CK spaces
$SO_{\k_1,\k_2}(4)/SO_{\k_2}(3)$ expressed in geodesic polar coordinates according  to
$\k_1,\k_2\in\{\pm 1,0\}$. }}
\label{table2}
\medskip
\noindent\hfill
$$
\begin{array}{lll}
\hline
\\ [-6pt]
\mbox {$\bullet$   Sphere $   {\bf S}^3$} &\quad\mbox {$\bullet$  Euclidean 
$ {\bf E}^3$  
 }&\quad\mbox {$\bullet$ Hyperbolic  $  {\bf H}^3$ } \\[2pt] 
SO(4)/SO(3) &\quad    ISO(3)/SO(3)
 &\quad       SO(3,1)/SO(3)  \\ 
(\k_1,\k_2)=(+1,+1)&\quad (\k_1,\k_2)=(0,+1)&\quad (\k_1,\k_2)=(-1,+1)   \\[2pt]
\displaystyle{\d s^2_{\rm CK} =   \d \rr^2    } &\quad
\displaystyle{\d s^2_{\rm CK} =\ \d \rr^2  }&\quad
\displaystyle{\d s^2_{\rm CK} =  \d \rr^2   } \\[2pt]
\displaystyle{\quad  +  {\sin^2 \rr } \left( 
\d \theta^2 +  {\sin^2  \theta }  \,\d\tes^2 \right) } &\quad
\displaystyle{\quad  + \rr^2 \left( 
\d \te^2 + {\sin^2 \te }  \,\d\tes^2 \right)  }&\quad
\displaystyle{\quad +\sinh^2  \rr   \left( 
\d \te^2 +  {\sin^2  \te } \,\d\tes^2 \right) } \\ 
 \displaystyle{K_{ij}= +1\quad K =+6 }&\quad
 \displaystyle{K_{ij}= 0\quad K =0 }  &\quad
 \displaystyle{K_{ij}= -1\quad K =-6} \\[8pt]
\mbox {$\bullet$  Oscillating NH ${\bf NH}^{2+1}_+$}&\quad\mbox {$\bullet$ Galilean ${\bf
G}^{2+1}$}&\quad\mbox {$\bullet$
Expanding NH ${\bf NH}^{2+1}_-$}\\[2pt] 
\mbox {$T_4\odot(SO(2)\otimes SO(2))/ISO(2) $}&\quad\mbox {$IISO(2)/ISO(2) $}&\quad\mbox
{$T_4\odot(SO(1,1)\otimes SO(2))/ISO(2) $}\\ 
(\k_1,\k_2)=(+1,0)&\quad
(\k_1,\k_2)=(0,0) &\quad (\k_1,\k_2)=(-1,0)   \\[2pt]
 \displaystyle{\d s^2_{\rm CK} =   \d \rr^2   
 } &\quad
 \displaystyle{\d s^2_{\rm CK} =   \d \rr^2   } &\quad
 \displaystyle{\d s^2_{\rm CK} =   \d \rr^2    } \\ 
 \displaystyle{K_{ij}= +1\quad K =+6 } &\quad
 \displaystyle{K_{ij}=0 \quad K =0} &\quad
 \displaystyle{K_{ij}=-1 \quad K =-6}\\[8pt]
\mbox {$\bullet$ Anti-de Sitter   ${\bf AdS}^{2+1}$ }&\quad\mbox {$\bullet$ Minkowskian   ${\bf
M}^{2+1}$}&\quad\mbox {$\bullet$ de Sitter  ${\bf dS}^{2+1}$}\\[2pt] 
  SO(2,2)/SO(2,1) &\quad   ISO(2,1)/SO(2,1)
 &\quad      SO(3,1)/SO(2,1)  \\  (\k_1,\k_2)=(+1,-1)&\quad
(\k_1,\k_2)=(0,-1) &\quad (\k_1,\k_2)=(-1,-1)   \\[2pt]
\displaystyle{\d s^2_{\rm CK} =  \d \rr^2   } &\quad
\displaystyle{\d s^2_{\rm CK} =  \d \rr^2   }  &\quad
\displaystyle{\d s^2_{\rm CK} =  \d \rr^2   }\\[2pt]
\displaystyle{\quad  -{\sin^2  \rr }  \left( 
\d \te^2 +  \sinh^2  \te \,\d\tes^2 \right)  } &\quad
\displaystyle{\quad -   \rr^2 \left( 
\d \te^2 +  \sinh^2  \te  \,\d\tes^2 \right) }  &\quad
\displaystyle{ \quad  - \sinh^2 \rr  \left( 
\d
 \te^2 +  \sinh^2  \te  \,\d\tes^2 \right) }\\[2pt]
\displaystyle{K_{ij}= +1\quad K =+6 } &\quad
 \displaystyle{K_{ij}=0 \quad K =0}  &\quad
 \displaystyle{K_{ij}=-1 \quad K =-6 } \\[4pt]
\hline
\end{array}
$$
\hfill}
\end{table}


\sect{Contraction, curvature  and quantum algebras}


\subsect{A quantum deformation of  $sl(2,\mathbb R)$}

Let us consider the algebra of formal power series in a {\em real} deformation parameter $z$
($q={\rm e}^z$) with coefficients in $U(sl(2,\mathbb R))$. If this algebra is endowed  with a
(deformed)   Hopf structure \cite{Abe} we get   the  so called  {\em non-standard quantum
deformation}  of
$U(sl(2,\mathbb R))$, here denoted  by $U_z(sl(2,\mathbb R))\equiv sl_z(2)$.  The Poisson analogue of
this quantum algebra is given by  the following deformed Poisson brackets and coproduct map
$\Delta$~\cite{plb,Ohn}: 
\begin{equation}
 \{J_3,J_+\}=2 J_+ \cosh z J_-  , \qquad  
\{J_3,J_-\}=-2\,\frac {\sinh zJ_-}{z} ,\qquad   
\{J_-,J_+\}=4 J_3   , 
\label{ba}
\end{equation}
\begin{equation}
\begin{array}{l}
\Delta(J_-)=  J_- \otimes 1+ 1\otimes J_- ,\\[1pt]
\Delta(J_l)=J_l \otimes {\rm e}^{z J_-} + {\rm e}^{-z J_-} \otimes J_l   ,\quad l=+,3.
\end{array}
\label{bb}
\end{equation} The deformed Casimir function for (\ref{ba}) reads
\begin{equation}  {\cal C}= \frac {\sinh zJ_-}{z} \,J_+ -J_3^2  . 
\label{bc}
\end{equation}  A  one-particle symplectic realization of (\ref{ba}) is given by~\cite{plb}
\begin{equation}
 J_-^{(1)}=q_1^2 ,\qquad   J_+^{(1)}=
  \frac {\sinh z q_1^2}{z q_1^2}\,   p_1^2  ,\qquad J_3^{(1)}=
\frac {\sinh z q_1^2}{z q_1^2 }\,    q_1 p_1  ,
\label{bd}
\end{equation} so that  ${\cal C}^{(1)}=0$. By starting from (\ref{bd}),  the coproduct (\ref{bb})
determines the corresponding two-particle  realization of (\ref{ba}), which is defined on
$sl_z(2)\otimes sl_z(2)$:
\be  
\begin{array}{l}
\displaystyle{  \jm^{(2)}=q_1^2+q_2^2 ,\qquad 
 \jp^{(2)}=
\left(   \frac {\sinh z q_1^2}{z q_1^2}\, {\rm e}^{z q_2^2} \right)  p_1^2  +
\left( \frac {\sinh z q_2^2}{z q_2^2}\, {\rm e}^{-z q_1^2}   \right) p_2^2 },\\[10pt]
\displaystyle{ \jj^{(2)}=
\left( \frac {\sinh z q_1^2}{z q_1^2 } \, {\rm e}^{z q_2^2} \right) q_1 p_1  +
\left( \frac {\sinh z q_2^2}{z q_2^2 }\,   {\rm e}^{-z q_1^2} \right)  q_2 p_2 }   .
\end{array}
\label{bbe}
\ee  Then   the two-particle Casimir  is given by
\bea && {\cal C}^{(2)} = \frac {\sinh zJ_-^{(2)}}{z} \,J_+^{(2)} -\left(J_3^{(2)}\right)^2\nonumber\\
&&\qquad =\left(  \frac {\sinh z q_1^2 }{z q_1^2 } \,
\frac {\sinh z q_2^2}{z q_2^2} \,{\rm e}^{ -z
  q_1^2} {\rm e}^{ z q_2^2}
 \right)\left({q_1}{p_2} - {q_2}{p_1}\right)^2 . 
\label{bbf}
\eea  
This procedure can be iterated to arbitrary $N$. In particular the 3-sites coproduct,  $\Delta^{(3)}
=(\Delta \otimes \mbox{id})\circ
\Delta =(\mbox{id}\otimes \Delta )\circ \Delta $,  gives rise to  a three-particle  symplectic
realization  of (\ref{ba})    defined on $sl_z(2)\otimes sl_z(2)\otimes sl_z(2)$:
\bea 
&& \jm^{(3)}=q_1^2+q_2^2+q_3^2\equiv \>q^2 , \nonumber \\  
&&\displaystyle{  \jj^{(3)}=\left( 
\frac {\sinh z q_1^2}{z q_1^2 }  \, {\rm e}^{z q_2^2} {\rm e}^{z q_3^2} \, \right) q_1 p_1 +
\left(  \frac {\sinh z q_2^2}{z q_2^2 } \,{\rm e}^{-z q_1^2}{\rm e}^{z q_3^2}  \, \right) q_2 p_2     
}\nonumber \\  
&&\displaystyle{   \qquad\qquad \qquad  +
\left(  \frac {\sinh z q_3^2}{z q_3^2 } \, {\rm e}^{-z q_1^2}{\rm e}^{-z q_2^2} \, \right) q_3 p_3    
},\label{be}\\  
&&\displaystyle{     \jp^{(3)}= 
\left( \frac {\sinh z q_1^2}{z q_1^2}\, {\rm e}^{z q_2^2}{\rm e}^{z q_3^2}     \right) p_1^2 }  + 
\left(   \frac {\sinh z q_2^2}{z q_2^2} \, {\rm e}^{-z q_1^2} {\rm e}^{z
 q_3^2}   \right)   p_2^2  + 
\left(   \frac {\sinh z q_3^2}{z q_3^2} \,{\rm e}^{-z q_1^2}  {\rm e}^{-z q_2^2}    \right) p_3^2  .
\nonumber  
\eea
It is immediate to check that these three functions fulfil the  commutation
rules (\ref{ba}) with respect to the canonical Poisson bracket 
\be
\{f,g\}=\sum_{i=1}^3\left(\frac{\partial f}{\partial q_i}
\frac{\partial g}{\partial p_i} -\frac{\partial g}{\partial q_i} 
\frac{\partial f}{\partial p_i}\right).  
\label{bff}
\ee
Likewise, the  three-particle Casimir function ${\cal C}^{(3)}$ can be straightforwardly obtained.

In this way we have obtained  a  (three-particle) {\em quantum deformation}, in a generic Hopf algebra
framework,  of the $sl(2,\mathbb R)$ Lie--Poisson algebra, understood as a more general
structure  that depends on the ``additional" quantum deformation parameter $z$. This means that  under the {\em classical limit}
$z\to 0$   (or
$q\to 1$)  the non-deformed Lie--Poisson brackets, Casimir,
  primitive coproduct ($\Delta(X)=X\otimes 1 + 1\otimes X$) and   symplectic realization of 
$sl(2,\mathbb R)$ are recovered, the latter being   
$J_-^{(3)}=\>q^2$,  $J_+^{(3)}=  \>p^2=\sum_{i=1}^3 p_i^2$, $J_3^{(3)}=\>q\cdot\>p=\sum_{i=1}^3 q_i p_i$.

In the sequel we will show that    this quantum deformation can be interpreted  as an
algebraic/geometric tool that introduces  a {\em non-constant curvature} in a formerly {\em flat}  3D
Euclidean space $ {\bf E}^3$, in such a manner that the deformation parameter $z$  governs the
(non-constant) curvature of the underlying space. The number of copies of $sl_z(2,\mathbb
R)$  is just the dimensionality of the space, and further iterations of the coproduct map
would lead to an $N$D construction.


\subsect{Riemannian and (non-)relativistic spaces of  non-constant  curvature}

An infinite family of 3D 
  {\em free} (kinetic energy) Hamiltonians ${\cal T}$~\cite{sigmaOrlando} can be constructed in terms
of the generators (\ref{be}) in the form:
\be  
\begin{array}{l} {\cal T}=\frac 12 \jp^{(3)}\, f (z\jm^{(3)} ),
\end{array}
\label{ahaa}
\ee  where $f$ is an arbitrary  smooth function such that $\lim_{z\to 0}f(zJ_-^{(3)})=1$;  in this
way $\lim_{z\to 0}{\cal T}=\frac 12 \>p^2$ gives the usual kinetic energy on   ${\bf E}^3$. Thus by
writing
 (\ref{ahaa})  as the free Lagrangian, 
\be
 {\cal T}=\frac 12 \left( \frac
 {z q_1^2}{\sinh z q_1^2} \, {\rm e}^{-z q_2^2}{\rm e}^{-z q_3^2} \dot q_1^2   +
 \frac {z q_2^2}{\sinh z q_2^2} \, {\rm e}^{z q_1^2}{\rm e}^{-z q_3^2}
\dot q_2^2    +
 \frac {z q_3^2}{\sinh z q_3^2} \, {\rm e}^{z q_1^2}{\rm e}^{z q_2^2}
\dot q_3^2  \right) f (z\>q^2 ),
 \label{ca}
\ee 
 we obtain the geodesic flow on the 3D
   space whose definite positive  metric is  given  by
\be
 \d s^2=\left( \frac {2z q_1^2}{\sinh z
 q_1^2} \, {\rm e}^{-z q_2^2}{\rm e}^{-z q_3^2} \,\d q_1^2   +
  \frac {2 z q_2^2}{\sinh z q_2^2} \, {\rm e}^{z q_1^2}{\rm e}^{-z q_3^2}\, \d q_2^2   +
  \frac {2 z q_3^2}{\sinh z q_3^2} \, {\rm e}^{z q_1^2}{\rm e}^{z q_2^2}\, \d q_3^2 \right) \frac{1}{f
(z\>q^2 )}.
 \label{ccd}
\ee If one computes the corresponding sectional $K_{ij}$ and scalar $K$  curvatures associated to the
metric (\ref{ccd}) one finds that, in general, these are non-constant; the latter turns out to be 
\be
 K(x)=z\left( 6 f^\prime(x)\cosh x  +\left(
  4  f^{\prime\prime}(x)-5f(x)-5{f^\prime}^2(x)/f(x)
 \right) \sinh x
 \right) ,
\label{co}
 \ee where  $x\equiv z\jm^{(3)} =z  \>q^2$,     $f^\prime(x)=\frac{{\rm d}f(x)}{{\rm d}
 x}$ and $f^{\prime\prime}(x)=\frac{{\rm d}^2f(x)}{{\rm d} x^2}$. This, in turn, means that   the
deformed  coalgebra process  can be understood as the introduction  of a non-constant curvature on a
formerly   flat space
${\bf E}^3$. Hence 
  the non-deformed or ``classical" limit $z\to 0$  can then be identified with a proper {\em flat
contraction}, under the which, the metric (\ref{ccd}) reduces to  the 3D Euclidean one   in Cartesian
coordinates,   $\d s^2=\sum_{i=1}^3 \d q_i^2$, and the scalar curvature (\ref{co}) vanishes for any
choice of the   arbitrary function $f$ (which always reduces to 1).

Furthermore the   metric (\ref{ccd}) can be rewritten in order to give rise to curved spaces of 
pseudo- and semi-Riemannian type as well (with Lorentzian and degenerate metrics), which can be
thought as non-constant curvature deformations of  the CK spaces described in section 2.3.
Explicitly, we  apply
 the following change of coordinates from $\>q$ to the polar-type ones $(r,\te,\tes)$ (compare to
(\ref{bbb}) for $N=3$):
\bea   &&
\cos^2(\la_1r)= {\rm e}^{-2z \>q^2}, \cr &&
\tan^2(\la_1r)\cos^2(\la_2\te)={\rm e}^{2 z q_1^2}{\rm e}^{2 z q_2^2}\bigl({\rm e}^{2 z q_3^2}-1 
\bigr) ,\cr &&
\tan^2(\la_1r)\sin^2(\la_2\te)\cos^2\tes={\rm e}^{2 z q_1^2}
\bigl({\rm e}^{2 z q_2^2}-1 \bigr),\label{xc} \\  &&
\tan^2(\la_1r)\sin^2(\la_2\te)\sin^2\tes= {\rm e}^{2 z q_1^2}-1 ,
\nonumber
\eea  where we have denoted $z=\la_1^2$ and we have introduced an additional parameter  $\la_2$     
which can  be either a real  or a pure imaginary number~\cite{plb}. In this way, we find that the
initial Riemannian metric (\ref{ccd}) is mapped into 
\bea && \d s^2= 
 \frac{1}{\cos(\la_1 \rr) \ff (\la_1 r )} \left( \d \rr^2  +\la_2^2\,\frac{\sin^2(\la_1 \rr)}{\la_1^2}
\left( 
\d
 \te^2 + \frac{\sin^2(\la_2 \te)}{\la_2^2} \,\d\tes^2  \right) \right)  \nonumber\\ &&\quad\ \ = 
\frac{1}{\cos(\la_1 \rr) \ff (\la_1 r )}\, \d s_{\rm CK}^2 ,
 \label{xd}
\eea where $\ff(\la_1 r )\equiv f(z\>q^2)$ is   an arbitrary  smooth function such that
$\lim_{\la_1\to 0}\ff(\la_1 r )=1$.  Thus we have obtained a family of metrics, parametrized by
$\la_1,\la_2$ and depending on   the function     $\ff$, which  is just the    metric  of  the 3D CK
spaces of constant curvature $\d s^2_{\rm CK}$ (\ref{bbc})   multiplied by a ``conformal" factor $ 
{1}/({\cos(\la_1
\rr)\ff (\la_1 r )})$, once we identify
\be
\k_1\equiv z=\la_1^2 ,\quad \k_2\equiv\la_2^2 ,\quad\k_3= +1.
\ee
 Consequently, $z$ plays a {\em threefold role} as a
quantum deformation/contraction/curvature parameter, while $\la_2$ is a (graded) classical 
contraction/signature parameter  which allows us to deal, simultaneously, with Riemannian,
Lorentzian and degenerate metrics.

 In this new coordinates the  scalar curvature $K$ (\ref{co}) reads
\be K(y)=2 z \cos y\left( \left( \frac{1+ 3 \cos^2y}{\sin 2 y }\right) g^\prime(y) + 
g^{\prime\prime}(y) -\frac 54
\,\frac{{g^\prime(y)}^2 }{g(y)} - \frac 54\, g(y) \, \tan^2y\right),
\label{xe}
\ee where the radial variable $y=\la_1 r$. Then, according to the real values that $z=\la_1^2$ and
$\la_2^2$ can take, we find that (\ref{xd}) comprises the following types of   spaces:

\begin{itemize}

\item[$\bullet$]  When  $z=\la_1^2>0$, we obtain   a  family of 3D ``deformed"
 spherical ${\bf S}^3_z$ $(\la_2^2>0)$, oscillating NH ${\bf NH}^{2+1}_{+,z}$
$(\la_2=0)$ and anti-de Sitter   ${\bf AdS}^{2+1}_z$   $(\la_2^2<0)$ spaces. For $z=1$  $(\la_1=1)$,
their scalar   curvature  
 reads
$$ K(r)=2   \cos r\left( \left( \frac{1+ 3 \cos^2r}{\sin 2 r }\right) g^\prime(r) + 
g^{\prime\prime}(r) -\frac 54
\,\frac{{g^\prime(r)}^2 }{g(r)} - \frac 54\, g(r) \, \tan^2r\right).
\label{xea}
$$

\item[$\bullet$] When  $z=\la_1^2=0$ we recover the proper {\em flat}   
 ${\bf E}^3$  $(\la_2^2>0)$,   ${\bf G}^{2+1}$  $(\la_2^2=0)$ and    ${\bf M}^{2+1}$   $(\la_2^2<0)$
spaces of table \ref{table2}, all of them  with   $K_{ij}=K=0$. The underlying symmetry remains as a
Lie--Poisson one (non-deformed).

\item[$\bullet$]  And when  $z=\la_1^2<0$, we get a family  of ``deformed" 3D   hyperbolic ${\bf
H}^3_z$
$(\la_2^2>0)$, expanding NH ${\bf NH}^{2+1}_{-,z}$ $(\la_2=0)$ and de Sitter   ${\bf dS}^{2+1}_z$  
$(\la_2^2<0)$ spaces, with scalar curvature for $z=-1$ $(\la_1={\rm i})$ given by 
$$ 
K(r)= - 2   \cosh r\left( \left( \frac{1+ 3 \cosh^2r }{{\rm i} \sinh 2 r }\right) g^\prime({\rm i} r ) +
g^{\prime\prime}({\rm i} r ) -\frac 54
\,\frac{{g^\prime({\rm i} r)}^2 }{g({\rm i} r)} + \frac 54\, g({\rm i} r) \, \tanh^2r\right) .
\label{xec}
$$

\end{itemize}

In order to  illustrate explicitly the above results we consider the
 simplest example     corresponding to set $ f (z\jm^{(3)} ) =f (z\>q^2 )= \ff(\la_1 r )\equiv 1$ in
(\ref{ccd}), that is, ${\cal T}=\frac 12 \jp^{(3)}$. In this case  the sectional $K_{ij}$ and scalar $K$
curvatures in the coordinates $\>q$ turn out to be~\cite{sigmaOrlando}:
 \bea  
&& K_{12}=\frac z4 \,{\rm e}^{-z \>q^2}\left( 1+ {\rm e}^{2 z q_3^2}- 2 {\rm e}^{2z \>q^2}\right) ,
\nonumber \\   
&& K_{13}=\frac z4
\,{\rm e}^{-z \>q^2}\left( 2- {\rm e}^{2 z q_3^2}+ {\rm e}^{2 z (q_2^2+q_3^2)}- 2 {\rm e}^{2z
\>q^2}\right)  ,  \label{bjk}\\   
&&K_{23}=\frac z4 \,{\rm e}^{-z \>q^2}\left( 2-  {\rm e}^{2 z( q_2^2+
q_3^2)}- 2 {\rm e}^{2z
\>q^2}\right)  ,   \nonumber \\  
&& K=2 (K_{12}+K_{13}+K_{23} ) = -5 z \sinh(z\>q^2) . 
\nonumber
\eea  
In the polar-type coordinates with metric $\d s^2= \d s_{\rm CK}^2 /\cos(\la_1 \rr)$
  these curvatures read
\be
  K_{12}=K_{13}= -\frac 12 \la_1^2 \,\frac{\sin^2(\la_1
\rr)}{\cos(\la_1
 \rr)}, \quad K_{23}=\frac{1}{2}K_{12},\quad K= -\frac 52 \la_1^2
\,\frac{\sin^2(\la_1 \rr)}{\cos(\la_1
 \rr)} .
\ee  
We display in table \ref{table3} the six particular ``deformed" spaces (with non-constant curvature)
arising for  $\ff=1$. We omit the Euclidean, Galilean and Minkowskian   spaces, with $z=0$, as these
remain flat/non-deformed  as given in table \ref{table2}.

We remark that, in general, other choices for the geodesic motion Hamiltonian  (\ref{ahaa}) (with
$f\ne 1$) give rise to more complicated spaces of  non-constant curvature. We also stress that the
nine CK spaces of table
\ref{table2} can also be directly recovered from an
$sl_z(2)$-coalgebra symmetry   by setting $\ff(\la_1 r ) =1/ \cos(\la_1 r)$    ($f(z\>q^2)={\rm
e}^{z\>q^2}$),   that is, ${\cal T}=\frac 12 \jp^{(3)} {\rm e}^{ z \jm^{(3)}}$. This is 
 a very singular case amongst the whole family of curved spaces determined by  the metric (\ref{xd})
since in this case all the curvatures are   {\em constant}:   
$K_{ij}=z\equiv \k_1$ and $K=6z\equiv 6\k_1$.
Once again, the role of the deformation parameter $z$ as a curvature becomes striking.

\begin{table}[t] {\footnotesize
 \noindent
\caption{{Metric, sectional  and scalar curvatures of   six 3D   spaces  of
non-constant curvature expressed in   polar-type coordinates  with 
 $z=\la_1^2\in\{\pm 1\}$ and $\la^2_2\in\{\pm 1,0\}$. For the six cases the sectional  curvature
$K_{23}= {K_{1j}}/2$  with
$j=2,3$.}}
\label{table3}
\medskip
\noindent\hfill
$$
\begin{array}{ll}
\hline
\\ [-6pt]
\mbox {$\bullet$ Deformed   sphere ${\bf S}^3_z$} &\quad\mbox {$\bullet$ Deformed hyperbolic   ${\bf
H}^3_z$ } \\[4pt]  z=+1;\ (\la_1,\la_2)=(1,1) &\quad z=-1;\  (\la_1,\la_2)=({\rm i},1)   \\[4pt]
\displaystyle{\d s^2 = \frac{1}{\cos\rr} \left(  \d \rr^2 +  {\sin^2 \rr } \left( 
\d \te^2 +  {\sin^2  \te }  \,\d\tes^2 \right)  \right)  }  &\quad
\displaystyle{\d s^2 = \frac{1}{\cosh\rr} \left(  \d \rr^2    +\sinh^2  \rr   \left( 
\d \te^2 +  {\sin^2  \te } \,\d\tes^2 \right) \right) } \\[6pt]
  \displaystyle{ K_{1j} = - \frac{\sin^2 
\rr}{2\cos 
 \rr}\quad K= - \frac{5\sin^2  \rr}{2\cos 
 \rr } } &\quad
 \displaystyle{ K_{1j} = -  \frac{\sinh^2
\rr}{2\cosh 
 \rr} \quad K= - 
 \frac{5\sinh^2  \rr}{2\cosh 
 \rr}} \\[12pt]
\mbox {$\bullet$ Deformed oscillating NH   ${\bf NH}^{2+1}_{+,z}$} &\quad\mbox {$\bullet$ Deformed
expanding NH ${\bf NH}^{2+1}_{-,z}$ }\\[4pt] z=+1;\ (\la_1,\la_2)=(1,0) &\quad z=-1;\ 
(\la_1,\la_2)=({\rm i},0)   \\[4pt]
 \displaystyle{\d s^2 =  \frac{1}{\cos\rr}  \,   \d \rr^2   
 }  &\quad
 \displaystyle{\d s^2 =   \frac{1}{\cosh\rr}  \,  \d \rr^2    } \\[6pt]
 \displaystyle{ K_{1j} = - \frac{\sin^2 
\rr}{2\cos 
 \rr} \quad K= - \frac{5\sin^2  \rr}{2\cos 
 \rr }}   &\quad
 \displaystyle{ K_{1j} = -  \frac{\sinh^2
\rr}{2\cosh 
 \rr} \quad K= - 
 \frac{5\sinh^2  \rr}{2\cosh 
 \rr}}\\[12pt]
\mbox {$\bullet$ Deformed anti-de Sitter   ${\bf AdS}^{2+1}_z$ } &\quad\mbox {$\bullet$ Deformed de
Sitter  ${\bf dS}^{2+1}_z$}\\[4pt] z=+1;\ (\la_1,\la_2)=(1,{\rm i}) &\quad z=-1;\  (\la_1,\la_2)=({\rm
i},{\rm i})  \\[4pt]
\displaystyle{\d s^2 =  \frac{1}{\cos\rr} \left(   \d \rr^2  -{\sin^2  \rr }  \left( 
\d \te^2 +  \sinh^2  \te \,\d\tes^2 \right)  \right) }   &\quad
\displaystyle{\d s^2 =   \frac{1}{\cosh\rr} \left(  \d \rr^2 - \sinh^2 \rr  \left( 
\d
 \te^2 +  \sinh^2  \te  \,\d\tes^2 \right) \right) }\\[6pt] 
\displaystyle{ K_{1j} = - \frac{\sin^2 
\rr}{2\cos 
 \rr} \quad K= - \frac{5\sin^2  \rr}{2\cos 
 \rr } }  &\quad
 \displaystyle{K_{1j} = -  \frac{\sinh^2
\rr}{2\cosh 
 \rr} \quad K= - 
 \frac{5\sinh^2  \rr}{2\cosh 
 \rr} } \\[8pt]
\hline
\end{array}
$$
\hfill}
\end{table}


\section{Concluding remarks}

The aim of this paper is to illustrate how a   curvature can be understood either as a
contraction parameter or as a quantum deformation one. This is explictly achieved by 
constructing, respectively, a family of  symmetrical homogeneous CK spaces from a
theoretical Lie group approach and some  non-constant curved spaces from a quantum group
one.  We remark that although the CK algebras/spaces have been already described in
arbitrary dimension $N$, their    quantum deformed counterpart has only been presented here
for $N=3$. We recall that the coalgebra procedure~\cite{sigmaOrlando,coalgebra}  affords
for the $N$D generalization of any 2D result which, in fact, comes from the coproduct of
the quantum algebra (so covering all the expressions given in subsection 3.1), but a clear
geometrical/physical interpretation of the non-constant curved spaces is not so
straightforward. A deeper study of the $N$D coalgebra curved spaces is currently under
investigation.

On the other hand, from a dynamical viewpoint,  all the geodesic motions associated to the
family of (quantum deformed) metrics (\ref{xd}) are, in general, superintegrable since    they are
endowed with {\em three} functionally independent integrals of motion, besides the free Hamiltonian.
Such integrals come from the 2- and 3-particle Casimirs, and can be explicitly constructed.
Nevertheless,  by using the  coalgebra approach  there is  always  a constant of the motion  left in
order to ensure   maximal superintegrability (this is a completely general fact~\cite{sigmaOrlando}).
Such a family of geodesic motion  Hamiltonians associated to (\ref{xd}), in coordinates
$(r,\te,\tes)$ and canonical conjugated momenta $(p_r,p_\te,p_\tes)$, reads
\be 
{T} =\frac 12 { \cos(\la_1 \rr) \ff (\la_1 r ) }
 \left( p_\rr^2  + \frac{\la_1^2}{\la_2^2\sin^2(\la_1 \rr)} \left(   
 p_\te^2 + \frac{\la_2^2}{\sin^2(\la_2 \te)}\,  p_\tes^2  \right)
\right) ,
 \label{ma}
\ee 
where ${T} = 2{\cal T} $ (\ref{ahaa}). In this respect, we also stress that   different
superintegrable  potentials~\cite{sigmaOrlando} on   curved spaces with
$sl_z(2)$-symmetry can be obtained by adding a potential term  
${U}(zJ_-)$ to   ${T}$ (\ref{ma}), since the superintegrability
properties of the complete Hamiltonian ${H}={T}+{U}$ can be shown to be preserved due  to the
underlying (quantum) coalgebra symmetry. 
 Moreover, for the particular CK  metrics (with $\ff(\la_1 r ) =1/ \cos(\la_1 r)$) it  is possible
obtain   the additional integral by Lie algebraic methods, so that the corresponding
kinetic energy on the CK spaces is, as it is well known, maximally superintegrable.   Finally we
point out a fact worthy of consideration: although  the underlying (deformed and CK) curved spaces
are always well defined for any value of
$\la_1$ and $\la_2$,    their corresponding metrics cannot be used in a  dynamical picture
for the Newtonian spaces with degenerate metrics since if
$\la_2\to 0$, then $T\to
\infty$.



\section*{Acknowledgments}

\noindent This work was partially supported  by the Ministerio de Educaci\'on y Ciencia   (Spain,
Projects  FIS2004-07913 and MTM2005-09183),  by the Junta de Castilla y Le\'on   (Spain, Project
VA013C05), and by the INFN-CICyT (Italy-Spain).


\end{document}